\def \ba {\begin{array}}
\def \ea {\end{array}}

\def \bea {\begin{eqnarray}}
\def \eea {\end{eqnarray}}

\def \be {\begin{equation}}
\def \ee {\end{equation}}

\def\ni{\noindent}
\def\nn{\nonumber}

\def\l[{\left[}
\def\r]{\right]}

\documentstyle[12pt,a4]{article}

\begin{document}

\begin{center} 
{\bf Space-time Structures from Critical Values in 2D Quantum Gravity}
\footnote{Work partially
supported by the DGICYT.}
\end{center}
\bigskip
\bigskip
\centerline{ {\it V. Aldaya\footnote{E-mail: valdaya@iaa.es} 
and J.L. Jaramillo\footnote{E-mail: jarama@iaa.es}} }
\bigskip

\begin{itemize}
\item {Instituto de Astrof\'{\i}sica de Andaluc\'{\i}a (CSIC), Apartado Postal
 3004, 18080 Granada, Spain.}
\item  {Instituto Carlos I de F\'\i sica Te\'orica y Computacional, Facultad
de Ciencias, Universidad de Granada, Campus de Fuentenueva, 
Granada 18002, Spain.} 
\end{itemize}

\bigskip
\begin{center}
{\bf Abstract}
\end{center}
\small

\begin{list}{}{\setlength{\leftmargin}{3pc}\setlength{\rightmargin}{3pc}}
\item 
A model for 2D Quantum Gravity is constructed out of the Virasoro group. To 
this end the quantization of the abstract Virasoro group is revisited. For 
the critical
values of the conformal anomaly $c$, some quantum operators ($SL(2,R)$ 
generators) lose their 
dynamical content (they are no longer conjugated operators). The 
notion of space-time itself in 2D gravity then arises as associated with this 
kinematical 
$SL(2,R)$ symmetry. An ensemble of different copies of AdS  do 
co-exist in this model with different weights, depending on their curvature 
(which is 
proportional to $\hbar^{2}$) and they are connected by gravity operators. 
This model suggests that, 
in general, quantum diffemorphisms should not be imposed as constraints to 
the theory, except for the classical limit. 
\end{list}

\normalsize

\vskip 1cm

\section{Introduction}

The Virasoro group has been used in previous approaches to 2D quantum gravity,
leading to the construction of the action functional of 2D Polyakov induced
gravity \cite{Polyakov} (gravitational Wess-Zumino-Witten action), 
\be
S=-\frac {1}{48\pi}(\int d^2x \frac {\partial_+ F}{\partial_- F}
(\frac {\partial_{-}^3 F}{\partial_- F} - 2 \frac{(\partial_{-}^2 F)^2}
{(\partial_{-} F)^2})) \hbox{ \ }\hbox{,}\hbox{ \ } 
ds^2=\partial_- Fdx^{-}dx^{+}. \label{Pol} \\
\ee
In \cite{Alekseev} a coadjoint orbit method was employed, while a Group 
Approach to Quantization (GAQ) \cite{Victor}  was the main tool
in \cite{Miguel}.

These approaches share the use of a particular realization of the
Virasoro group as the central extension of  $diffS^1$ (i.e.
$di\tilde{ff}S^1$). Therefore, the space(-time) sub-manifold $S^1$ appears
in these constructions in an explicit way from the very beginning  and 
$\sigma$, the parameter of such a space, emerges as a variable external 
to the group. Another external evolution parameter
$t$, the domain of which is not well understood, then enters the theory
constituting 
with $\sigma$ the space-time manifold.

In our present framework, the Virasoro
group is taken firstly as an abstract group and secondly as the only physical
input 
of the theory. In particular, we do not assume the existence {\it a priori} of
a space-time on which a quantum field theory (of gravitation) is constructed, 
and this entity must be one of the 
results of our quantization process of a fundamental symmetry group. In 
fact, the structure of space-time is one of our main 
objectives.

The method of quantization that we use is GAQ again, although
paying special attention to its quantum aspects. We briefly present the 
fundamentals of such a technique (for more precise
details the reader can refer to \cite{Victor}).

The chief aim of GAQ is the construction of an unitary and irreducible
representation of a Lie group $G$ of physical operators. The operators
in this algebra are classified into two classes: those providing a
central  term under commutation, which are the dynamical ones, and the
remainder or the kinematical ones. The former ones appear in
conjugated pairs and are the basic variables, while the latter ones
generate transformations on the former and are eventually written in
terms of them. Therefore the presence of  central  terms is essential
for the existence  of dynamics. If they do not appear, then the
algebra must be  (pseudo-)extended, provided that its
(pseudo-)cohomology group allows this \cite{julio}. Indeed, the
treatment of anomalies in this context \cite{anomalias} advises us to
extend maximally the original algebra.

The extended group  $\tilde{G}$ has the structure of a principal fibre
bundle, with fibre $U(1)$ over $G$, with connection 1-form $\Theta$, 
selected as a component of the 
canonical left-invariant 1-form dual to the fundamental (or vertical) vector
$\Xi$.
The characteristic subalgebra of $\Theta$, denoted by ${\cal G}_{\Theta}$,
generates the characteristic module of $\Theta$ -that is, $Ker\Theta\cap Ker
\hbox{d}\Theta$- and
coincides with the kernel of the Lie-algebra cocycle. It thus contains
the non-dynamical generators.

The {\it prequantization} is accomplished by the regular representation (over
the complex $U(1)$-functions defined on the group). This provides us with two
copies
of generators: left-invariant ($\tilde{X}^L$) and right-invariant ($\tilde
{X}^R$) vector fields. The main
point is that these two sets of generators do commute, and this allows us to 
consider one set as the physical operators (namely the right-invariant ones), 
and to use 
the others (left-invariant ones) to find a 
polarization to reduce the representation, thus arriving to the true  
{\it quantization} (obviously, we can interchange the 
role of the left- and right-invariant vector fields, while maintaining the 
same physical system).

An important feature of the present approach is that no use of Lagrangians, 
actions, etc., is required
to formulate the physical theory. Those objects are of no primary interest to
us. Only the Hilbert space and the action of quantum operators on it have 
such a fundamental meaning. However, GAQ also provides a generalization of
the Cartan approach to Mechanics, which is the natural framework in which to 
discuss the
classical aspects of the theory. In fact, the 1-form $\Theta$ is the 
generalization of the Poincar\'e-Cartan form and the trajectories of the 
(left-invariant) vector fields in ${\cal G}_{\Theta}$ constitute the 
generalized classical equations of motion. The classical solution manifold
$\tilde{G}/({G}_\Theta\otimes U(1))$
(phase space) is parametrized by the Noether invariants; that is, the functions
$i_{\tilde{X}^R}\Theta$, which are constant along the classical trajectories 
($L_{\tilde{X}_j}i_{\tilde{X}^R}\Theta=0,\; \forall\; \tilde{X}_j\;\in\;
{\cal G}_\Theta$). The presymplectic form $\hbox{d}\Theta$ falls down to the
phase space defining a true symplectic form and the corresponding Poisson 
bracket. The action is 
obtained from $\Theta$ by integrating it along certain trajectories 
$\sigma(\tau)$ on the group $\tilde{G}$.

We emphasize the fact that physical systems are quantum mechanical so that
we shall use the previous classical machinary only after the classical limit
to identify the underlying physics. This classical identification is of 
crucial relevance for an abstract approach to quantum theory such as the 
present one (see section IV in the first reference of \cite{Victor}).

Once the fundamentals of our technical background are presented,
we refine somewhat more the main question; that is, whether the space-time 
concept itself emerges from our treatment and, if so, how.
The answer is that the space-time variables must be sought as related
to the operators inside the characteristic subalgebra ${\cal G}_\Theta$.

In a classical approach to field theory, the space-time variables are the 
integration parameters of certain 
generators inside the subalgebra ${\cal G}_\Theta$ which, as stated above, 
generate movements 
in phase-space variables (as space-time translations do). Along the 
corresponding trajectories, the dynamical parameters
in the group gain a dependence in these integration parameters, thus 
becoming fields over them.
In fact, in the process of obtaining the classical action functional out of
the field $\Theta$, we can identify the space-time variables, after 
solving the equations of motion for the generators in ${\cal G}_{\Theta}$,
as those appearing explicitly in the integration measure. This construction of
the space-time support from the group, can be explicitly
shown in the case of Poincar\'e invariant dynamics for the
scalar, electromagnetic and Proca fields. In these cases, we can begin from
the corresponding groups (see \cite{empro} and references there in), without 
considering the space-time
and reconstruct it after the exact resolution of the motion equations. The 
kinematical symmetry group proves to be 
contained in the fields group.

However, the natural way of approaching to the space-time underlying a
quantum (field) theory would consist of finding the support for the
quantum states of the  irreducible Hilbert space of the theory,
through the $C^*$-algebra defined by those states.

\section{The Virasoro group}

\ni In this section, we present a quick survey of the Virasoro  group,
 our starting point being the algebra:
\be
\left[ \hat{L}_n,\hat{L}_m\right] =\left( n-m\right) \hat{L}_{n+m} \;\; . 
\label{Virasoronoex}
\ee

\ni As stated in the {\bf Introduction}, we shall consider all the central 
extensions of this algebra, which will decide the dynamical content of
the group parameters. Such extensions are:
\be
[ \hat{L}_n,\hat{L}_m] =( n-m) \hat{L}_{n+m}+\frac 1{12}( cn^3-c'n)
\delta _{n,-m} \;\; ,     \label{conmutacion}
\ee
\ni where $c$ is the genuine central extension parameter and $c'$ is the 
parameter of a family of pseudo-extensions (a redefinition of $L_0$
causing a non-trivial connection form on the group; see
\cite{Navarro-Salas}).

The next step is to construct a formal group law from this algebra,
and this was indeed done in \cite{Navarro-Salas}.  The resulting
expression for the extended group is \footnote{Throughout the text,
summation symbols are explicited only in those cases which present a
constraint on the indices. In all other cases, wherever an index
appears repeated, summation from $-\infty$ to $\infty$ is understood.}:
\bea 
l^{\prime\prime m}&=&l^{m}+l'^{m}+ipl'^{p}l^{m-p}+
\frac{(ip)^2}{2!}l'^{p}l^{n}l^{m-n-p}+...+
\sum\limits_{ n_1+...+n_j=-k}\frac{(ip)^r}{r!}l'^{p}
l^{n_1}...l^{n_r}+...\nn \\ 
\varphi ^{\prime \prime }&=&\varphi ^{\prime }+\varphi +\xi _c(
g,g') -\frac{c^{\prime }}{24}\xi _{cob}( g^{\prime },g) \;  \\ &&
\hbox{ \ \ \ } \xi _{cob}( g^{\prime },g)=l{{}^0}^{\prime \prime
}-l{{}^0} ^{\prime }-l{{}^0} \nn 
\eea

\ni (the explicit expression for $\xi _c\left(
g,g'\right)$ is rather involved and thus we refer the reader to
\cite{Navarro-Salas}).

From this group law, we compute the left- and  right-invariant  vector
fields, $\tilde{X}_{l^k}^L$ and $\tilde{X}_{l^k}^R$, respectively. The
corresponding expressions are  presented in {\bf Appendix B}. We make
explicit here only the non-central part of $\tilde{X}_{l^k}^L$:\\
\be
{X}_{l^k}^L=\frac \partial {\partial l^k}+i\left( m-k\right)
l^{m-k}\frac \partial {\partial l^m} \;\; . \label{xl}
\ee 

\ni The quantization form is obtained by duality on left
fields ( $\Theta( \Xi) =1,\;\;\Theta ( \tilde{X}_{l^n}^L) =0$):
\bea
\Theta&=& \frac{i}{24}(cn^2-c')nl^{-n}dl^n+ \nn \\
&+&\sum_{\stackrel{k=2}{n_1+...n_k=-n}}
\frac{(-i)^k}{24}[cn_1^2-c'+ 
cn^2\sum_{m=2}^k\frac{1}{m!}]n_1...n_kl^{n_1}...l^{n_k}dl^n +d\varphi 
\eea

Especially important in searching for the space-time notion is the
structure of the characteristic subalgebra ${\cal G}_\Theta $ of
$\Theta $,  which coincides with the kernel of the Lie algebra
cocycle. Thus, depending on the values of $c$ and $c^{\prime}$, we
find:\\
\bea
\hbox{i)} \frac{c^{\prime }}{c}\neq r^{2} , r\in Z, \; \Rightarrow \;
 {\cal G}_\Theta &=& \langle \tilde{X}_{l^0}^L\rangle \label{r} \\
\hbox{ii)} \frac{c^{\prime }}{c}=r^2, r\in Z, \; \Rightarrow \; 
{\cal G}_\Theta&=&\langle
\tilde{X}_{l^{-r}}^L,\tilde{X}_{l^o}^L,\tilde{X}_{l^r}^L\rangle
\approx sl^{( r) }( 2,R) \;\; . \nn
\eea

\ni Since we wish to find a two-dimensional space-time inside the group, we 
must choose $ii)$. Besides, we are searching for a unitary
representation of  our algebra.  This imposes (see
\cite{Navarro-Salas,Witten} and the next section)  $c=c^{\prime}
\;\; \;(r=1)$. We must note, however, that, although we need $c=c^{\prime}$ 
for implementing a 
notion of space-time, the dynamics of our system are as well defined
for other values of $c$ and $c^{\prime}$ (provided that the theory is
unitary) but without a notion of space-time as such. Our first
conclusion then is that  space-time appears as a critical case  and
outside this critical value of the conformal anomaly, we would still
have a well-defined physical system.

Let us now detail the elements of the Cartan-like geometry associated
with the Virasoro group in the critical case $c=c'$, which will
constitute the mathematical framework of the physical theory
underlying the classical limit.

We write the evolution equations for the $l^{n}$ parameters under the
action of $SL(2,R)$.  Under this action, $l^{n}$ are functions of the
$SL(2,R)$ parameters, thus becoming, as we stated in the  {\bf
Introduction}, fields over the $SL(2,R)$ manifold (parametrized by
$\tilde{\lambda}_0, \tilde{\lambda}_1, \tilde{\lambda}_{-1}$ which are
the  parameters associated to the fields
$\tilde{X}_{l^0}^{L},\tilde{X}_{l^1}^{L},
\tilde{X}_{l^{-1}}^L $, respectively):$l^n=l^n(\tilde{\lambda}_0, 
\tilde{\lambda}_1, \tilde{\lambda}_{-1})$.
The dynamical system can then be written out as:
\be
\frac {\partial l^n} {\partial \tilde{\lambda}_0}=(\tilde{X}_{l^0}^{L})^{l^n}
\hbox{\ \ , \ \ }\frac {\partial l^n} {\partial \tilde{\lambda}_1}=
(\tilde{X}_{l^1}^{L})^{l^n}\hbox{\ \ , \ \ }\frac {\partial l^n} {\partial
\tilde{\lambda}_{-1}}=(\tilde{X}_{l^{-1}}^{L})^{l^n},
\ee
\ni (where the $l^n$ component of the field $\tilde{X}_{l^i}^{L}$, 
$(\tilde{X}_{l^i}^{L})^{l^n}$, proves to be $\tilde{X}_{l^i}^{L}(l^n)$).

Using the explicit expressions for $(\tilde{X}_{l^m}^{L})^{l^n}$ from 
(\ref{xl}), we find:\\
\bea
\frac {\partial l^m} {\partial \tilde{\lambda}_0}&=&im l^{m}\hbox{\ \ \ for
 \ \ \ } 
 m\neq0 
\hbox {\ \ \ , \ \ \ }
\frac {\partial l^0} {\partial \tilde{\lambda}_0}=1 \nn \\
\frac {\partial l^m} {\partial \tilde{\lambda}_1}&=&
i(m-1)l^{m-1}\hbox{\ \ \ for \ \ \ } 
 m\neq1 
\hbox {\ \ \ , \ \ \ }
\frac {\partial l^1} {\partial\tilde{\lambda}_1}=1 \label{motion}\\
\frac {\partial l^m} {\partial\tilde{\lambda}_{-1}}&=&
i(m+1)l^{m+1}  \hbox{\ \ \ for \ \ \ } 
 m\neq-1 
\hbox {\ \ \ , \ \ \ }
\frac {\partial l^{-1}} {\partial\tilde{\lambda}_{-1}}=1 \hbox{ .} \nn 
\eea

The solutions to these equations can be obtained exactly:
\bea
l_0&=& \lambda_0 \nn \\
l_{-1}&=& [\lambda_{-1}+\sum_{s=0}^{\infty} (-1)^s
\left(\begin{array}{c} 1+s \\ 1 \end{array} \right)(i\lambda_{1})^s
 {\cal L}_{-1-s}]
e^{-i\lambda_0} \nn \\
l_1&=& [\lambda_1+\sum_{s=0}^{\infty} (-1)^s
\left(\begin{array}{c} 1+s \\ 1 \end{array} \right)(-i\lambda_{-1})^s
 {\cal L}_{1+s}]
 e^{i\lambda_0} \label{sol}  \\
l_{-n}&=&[\sum_{s=0}^\infty \sum_{m=0}^{n-2} [(-i)^{s+m}\left(\begin{array}
{c} n-1 \\ m \end{array} \right)\left(\begin{array}{c} s+n-m 
\\ n-m \end{array} \right)
\lambda_1^s\lambda_{-1}^m{\cal L}_{-n+m-s}] +\frac{i}{n}
(-i\lambda_{-1})^n]e^{-in\lambda_0} \nn \\
l_n&=& [\sum_{s=0}^\infty \sum_{m=0}^{n-2} [i^{s+m}\left(\begin{array}{c} n-1 
\\ m \end{array} \right)\left(\begin{array}{c} s+n-m \\ n-m \end{array} \right)
\lambda_{-1}^s\lambda_1^m{\cal L}_{n-m+s}] -\frac{i}{n}
(i\lambda_1)^n]e^{in\lambda_0} 
\;\;,n\geq 2 \;,\nn 
\eea
\ni where $\tilde{\lambda}_0= \lambda_0$, $\tilde{\lambda}_1=\lambda_1
e^{i\lambda_0}$, $\tilde
{\lambda}_{-1}=\lambda_{-1}e^{-i\lambda_0}$, and the integration constants 
${\cal L}_ n\; (\mid\! n\!\mid\geq 2)$, parametrize the solution manifold 
$Virasoro/SL(2,R)$.

This symplectic manifold, with symplectic form $d\Theta/SL(2,R)$,
can also be parametrized by the basic Noether invariants $L_n\equiv 
i_{X^L_{l^n}}\Theta$, with $\mid\! n\!\mid\geq 2$. The Noether invariants 
$L_j\equiv i_{X^L_{l^{j}}}\Theta\;(j=0,\pm 1)$ must be written in terms
of the basic ones. We shall illustrate this fact to the lowest non-trivial
order, at which:
\bea
L_k&=&i_{X^L_{l^k}}\Theta=\frac{ic}{12}(k^2-1)kl^{-k}-\frac {c}{24}
\sum_{n_1+n_2=-k}[(n_1^2-1+\frac{k^2}{2})n_1n_2 \nn \\
&-&(n_1^2-1)n_1k+\frac{k^2}{2}(n_1^2+n_2^2+n_1n_2)-\frac{k^2}{2}]l^{n_1}
l^{n_2}+... 
\eea
From this, we explicitly see that for the kinematical Noether invariants,
($L_0, L_{\pm 1}$), the linear term vanishes and the only contribution to the
cuadratic term comes from the basic Noether invariants \footnote{For $L_0$, 
$L_1$ and 
$L_{-1}$, the polynomial on $n_1$ and $n_2$ in the quadratic term vanishes 
whenever $l^0,l^1$ or $l^{-1}$ appear.}, as should be the case. For these
basic Noether invariants, we find $l^n=(\frac{12}{ic(k^2-1)k} L_{-n}+...)\;
,\mid\! n\!\mid\geq 2\;$, and the expressions of the kinematical ones are:
\bea
L_{0}&=&\frac{6}{c}\sum_{n\neq 0,\pm1}\frac 1{(n^2-1)}
L_{-n}L_{n}+... \label{invNoet}\\
L_{1}&=&\frac{6}{c}\sum_{n\neq 0,\pm1,-2}\frac 1{n(n+1)}
L_{-n} L_{n+1}+... \nn \\
L_{-1}&=&\frac{6}{c}\sum_{n\neq 0,\pm 1,2}\frac1{n(n-1)}L_{-n}L_{n-1}+... \nn 
\eea

The previous parametrization of the solution manifold with $L_n\;({\cal L}_n)$ 
or, accordingly, of the Virasoro group with $l_n$ corresponds to a 
Fourier-like description. A 
configuration-like description will be achieved by defining the field
\bea
F(\lambda_{-1},\lambda_0,\lambda_1)=\sum_n l_n(\lambda_{-1},\lambda_0,
\lambda_1)\;, 
\eea
\ni which parallels the standard Fourier expansion of a field, $\phi(x,t)
=\sum_k A_ke^{ikx-k_0t}$, where the constants ${\cal L}_n$ in (\ref{sol}) play
the role of the $A_k$'s, and the 
functions of $\lambda_0,\lambda_1,\lambda_{-1}$ accompanying the 
${\cal L}_n$'s play the role of the exponentials.  

Explicitly, and with some abuse of the language concerning the notation 
of $l_n(\lambda_{-1},\lambda_0,\lambda_1)$ and  $l_n(\lambda_{-1},
\lambda_1)$, 
\bea
& &F(\lambda_{-1},\lambda_0,\lambda_1)=\sum_n l_n(\lambda_{-1},\lambda_0,
\lambda_1)=   \label{F}  \\
&=&\lambda_0 + [\sum_{n>0}\sum_{s=0}^\infty \sum_{m=0}^{n-2} 
i^{s+m}\left(\begin{array}{c} n-1 \\ m \end{array} \right)\left
(\begin{array}{c} s+n-m \\ n-m \end{array} \right)
\lambda_{-1}^s\lambda_1^m{\cal L}_{n-m+s} -\frac{i}{n}(i\lambda_1)^n]
e^{in\lambda_0} 
+    \nn \\
&+& \sum_{n>0}\sum_{s=0}^\infty \sum_{m=0}^{n-2} (-i)^{s+m}\left
(\begin{array}{c} n-1 \\ m \end{array} \right)\left(\begin{array}{c} s+n-m \\
 n-m \end{array} \right)
\lambda_1^s\lambda_{-1}^m{\cal L}_{-n+m-s} +\frac{i}{n}(-i\lambda_{-1})^n]
e^{-in\lambda_0} = \nn \\
&=&\lambda_0+\sum_{n\neq0}l_n(\lambda_{-1},\lambda_1)e^{in\lambda_0} \nn
\eea
Making the change of variables:
\bea
u&=&\frac1{2}(\lambda_1+\lambda_{-1}) \nn \\
v&=&\frac1{2}(\lambda_1-\lambda_{-1})  \\
\lambda&=&\lambda_0 \;, \nn
\eea
\ni we express
\bea
F(u,v,\lambda)&=&\lambda+\sum_{n\neq0}l_n(u,v)e^{in\lambda}\;. 
\eea

As we shall see in the next section, the space-time notion is related to 
that of homogeneous spaces inside $SL(2,R)$. Both de Sitter and anti-de Sitter
are found among these spaces, since dS and AdS groups in two dimensions are
isomorphous to $SL(2,R)$. AdS geometry can be constructed from the Killing
metric,
\be
ds^2={dv}^2+{d\bar{\lambda}}^2-{du}^2 \;\; , \\
\ee
by imposing the Casimir constraint:\\
\be
{v}^2+{\bar{\lambda}}^2-{u}^2={R}^2 \;\; , \\
\ee
\ni where $\bar{\lambda}$ is the decompactified $\lambda$ ($\bar{\lambda}=
sin^{-1}\lambda$).

\ni The geometry of dS follows from:
\be
ds^2={dv}^2-{d\bar{\lambda}}^2+{du}^2 \;\; , \\
\ee
with
\be
{v}^2-{\bar{\lambda}}^2+{u}^2={R}^2 \;\; . \\
\ee
We see that they are topologically the same (a one-fold hyperboloid), but
AdS has negative constant curvature, $K=-\frac{1}{R^2}$, and compact time, 
while dS has positive constant curvature, $K=\frac{1}{R^2}$, and compact 
space. In both cases, Minkowski is recovered within the limit 
$R^2\rightarrow\infty$ \footnote{Constant curvature space-times in two 
dimensions are not a solution to the
Einstein field equations with cosmological constant $\Lambda$ ($\neq 0$) in 
a vacuum. In higher dimensions, however, they are, and we find 
$K\approx\Lambda$
\cite{Weinberg,Hawking}. Thus, in the hope that these results can be extended 
to 
higher dimensions in a suitable generalization, we are tempted to interpret
$K$ as a cosmological constant.}. 
\\ 
\\

\section{Quantum representations: a  model for the quantum theory of gravity}
\subsection{Algebraic construction}
Let us return to the problem of obtaining a unitary, irreducible 
representation of the Virasoro group. As stated above, this problem 
was studied in 
\cite{Kac,Friedan,Witten,Navarro-Salas} and we take the results from
 \cite{Navarro-Salas}.

Two cases are of interest to us: $\frac{c^{\prime }}{c}\neq r^{2}$ and
$\frac{c^{\prime }}{c}=1$. For both, we can find a full (including the 
entire characteristic 
subalgebra) and symplectic (including one of the two coordinates of each 
dynamical pair) polarization:\\
\bea
{\cal P}&=&\left\langle \tilde{X}_{l^{n\leq 0}}^L\right\rangle \hbox{\ \ \ 
for \ }\frac{c^{\prime }}{c}\neq r^{2} \\
{\cal P}^{\left( 1\right) }&=&\left\langle \tilde{X}_{l^{n\leq
1}}^L\right\rangle \hbox{\ \ \ for \ } c=c^{\prime} \hbox{ ,}\nn 
\eea
\ni and the corresponding polarization conditions for the wave 
functions $\Psi$:\\
\bea
\tilde{X}_{l^{n\leq 0}}^L\;\;\Psi =0\;\;\;\;\;&\hbox{if}&\;\;\;c\neq 
c^{\prime}\label{Polapsi} \;\; (\tilde{X}_{l^{n}}^L\in{\cal P})\\
\tilde{X}_{l^{n\leq 1}}^L\;\;\Psi =0\;\;\;\;\;&\hbox{if}&\;\;\;
c=c^{\prime }\;\; (\tilde{X}_{l^{n}}^L\in{\cal P}^{(1)}) \hbox{ .} \nn
\eea
\ni (Note that we can work with the case $c\neq c^{\prime }$ only because 
the latter can be
formally recovered from the former by making $c=c^{\prime}$ at the end of the 
calculations).

The solutions to these polarization equations build the representation Hilbert 
space. The Virasoro algebra operators are represented by acting
with the right-invariant vector fields on these specific polarized functions.

Redefining the generators:\\
\be
\hat{L}_{n\neq 0}=i\tilde{X}_{l^n}^R\;\;\;,\;\;\;\hat{L}
_0\equiv i \tilde{X}_{l^0}^R \;\;\;,\;\;\;I
\equiv i\Xi \;,\nn
\ee
\ni we recover the usual commutators for the 
Virasoro Lie algebra (these relationships are more usually expressed 
in terms of
$(c,h)$, where $h=\frac{c-c'}{24}$, but we prefer to maintain the $(c,c')$
parameters in which our analysis is more transparent).

It should be pointed out that in these representations there are no null
 vectors
\cite{Navarro-Salas}. This is a crucial point, because the space of polarized 
functions is not irreducible in general (a difference with the compact 
semisimple group case). Taking advantage of the absence of null vectors, 
it is possible to
consider the orbit of the enveloping algebra through the vacuum and thus to  
construct an irreducible subspace ${\cal H}_{( c,c') }$:\\
\bea 
{\cal H}_{\left( c,c'\right) }\equiv \left\langle \hat{L}
_{n_j}...\hat{L}_{n_1}\;\mid\! 0\rangle
\right\rangle \;\;\;\;\;n&\leq&-1\;\;\;j=1,2,3,...\;\hbox{ for }c\neq c'
\label{Hilbert}\\
n&\leq&-2\;\;\;j=1,2,3,...\; \hbox{ for }c=c'.
\eea
These are the representation spaces we shall work with.\\

With regard to unitarity and irreducibility, some brief comments are 
relevant:\\
\begin{itemize}
\item Values of $c$ and $c'$ for unitary representations:\\
{\it a)} $c \geq 1 \hbox{  , with  } \frac {c-c'}{24} \geq 0$.\\
{\it b)} $0<c<1$  with:
\bea
c&=& 1- \frac 6{m(m+1)} \\
\frac{c-c'}{24}&=& \frac {{[(m+1)r - ms]}^2-1}{4m(m+1)} \; \; \; (1\leq s\leq r
\leq m-1 \;,\;
\hbox{ and } m,r,s \hbox{ integers} \nn \\
\hbox{with } m\geq 2) 
\eea
\ni Pairs $(c,c')$  different from the previous ones, lead to non-unitary 
representations.
\item Values of $c$ and $c'$ for reducible representations:\\
\bea
c&=& 1- \frac 6{m(m+1)} \label{irred} \\
\frac{c-c'}{24}&=& \frac {{[(m+1)r - ms]}^2-1}{4m(m+1)} \; \; \; 
(r,s \in Z^+\;,\; m \in R^+)\hbox{.} \nn
\eea
For $c>1$, therefore, we have irreducible representations.
\end{itemize}

For more details about unitarity and irreducibility see 
\cite{Kac,Feigin&Fuchs,Thorn,Senechal,Itzykson,Navarro-Salas}. In particular, 
in the second reference
in \cite{Navarro-Salas}, it is proven that the reduction for $c<1$ can be 
achieved by means of higher-order polarizations.

The representation of our original algebra on a Hilbert space has been 
accomplished. As noted above, making $c=c^{\prime}$ (that is $h=0$), 
the Virasoro representations
with $SL^{(1) }( 2,R)$ as the 
characteristic subalgebra, are recovered. This is the
case in which a notion of space-time can be found. Under this condition, 
there are two kinds of operators acting on our Hilbert space:
\begin{itemize}
\item Dynamical operators ({\it gravity} field operators): $\hat{L}_{n}\;
\hbox{,}\; \mid n\mid\geq {2}$\\
\item Space-time operators: $\hat{L}_{n}\;\hbox{,}\; \mid n\mid\leq {1}$ . 
\end{itemize}

As a preliminary approach to the construction of an explicit model for Quantum
Gravity problem and, in order to simplify the mathematical issues related with
space-time reconstruction, we are going to focus on the case $c>1$. This 
condition, together with $c=c'\Leftrightarrow h=0$, guarantees unitarity,
irreducibility and allows for the notion of space-time. Although
there are unitary representations with $c=c'$ and $c\leq 1$ (with $r=s$ and 
thus parametrized by $m\geq 2$), these representations are reducible 
and we must resort to higher-order polarizations which lead to a 
non-commutative structure on the $C^*$-algebra of the functions 
in the carrier subspace. This problem, although extremely interesting, is 
beyond the scope of this work.

To begin the study of implementing of the space-time notion, 
let us consider the reduction of our unitary 
irreducible representation of the Virasoro group under its space-time subgroup 
$SL^{\left( 1\right) }\left(2,R\right)$. From the orbit-through-the-vacuum 
construction for the 
representation of the Virasoro group, the $SL^{(1)}(2,R)$ representations
(which are unitary and thus infinite-dimensional) are of maximal-weight type
(see
\cite{Lang}). As can be seen in detail in the {\bf Appendix A}, on each level 
of the
Virasoro representation (that is, the finite-dimensional space of eigenvectors
of $L_0$ with eigenvalue $N$) there exist $(D^{N}-D^{N-1})$  maximal weight 
vectors of $SL^{(1)}(2,R)$, where $D^{N}$ is the dimension of the $N$ level,
given by the number of partitions of $N$ in which $1$ is lacking (for 
instance, for $N=4$, $(2,2)$ is allowed while $(3,1)$ is not).
From each of these maximal-weight vectors, an irreducible representation
of $SL^{(1)}(2,R)$ with index $N$, $R^{(N)}$, and with Casimir $N(N-1)$, 
is constructed.

The reduction of the original Hilbert space, ${\cal H}_{(c,c)}$, is then:\\
\be
{\cal H}_{(c,c)}=\bigoplus_{N} (D^{(N)}-D^{(N-1)})R^{(N)}\hbox{.} 
\ee
\ni It can be shown that these $SL(2,R)$ irreducible representations are 
orthogonal with the Virasoro scalar product ($\hat{L}_n={\hat{L}_{-n}}^+$, 
$\langle 0\!\mid 
\!0\rangle =1$), allowing a standard quantum interpretation of the states.  
We note that $(D^{(N)}-D^{(N-1)})$, the degeneration of the $R^{(N)}$ 
representation, increases with $N$. 

We give examples of the $SL^{(1)}(2,R)$ representations with the lowest values
for $N$.
To do so, we look for $SL^{(1)}(2,R)$ maximal-weight vectors at level $N$
by considering the most general linear combination of Virasoro states
of level $N$ and then simply determining the coefficients for which this 
vector 
is annihilated by $\hat{L}_1$ (there are $(D^{(N)}-D^{(N-1)})$ solutions: the
kernel of $\hat{L}_1$ restricted to level $N$). 
The excited $SL^{(1)}(2,R)$ states are established by applying the operator 
$\hat{L}_{-1}$ successively on the corresponding vacuum.

For $N=1$, there are no Virasoro states (because $c=c'$), and thus there are
no $(N=1)-SL(2,R)$ representations.

For $N=2$, we have only the vector $\hat{L}_{-2}\mid\!0\rangle$, 
\footnote{$\mid\!0\rangle$ is the Virasoro vacuum.} which is in fact 
annihilated by $\hat{L}_1$ (as it should be). The excited states are:
\be
\mid N=2,n\rangle=C_{2,n}
(\hat{L}_{-1})^n\hat{L}_{-2}\mid0\rangle ,  
\ee
where $C_{2,n}$ is a normalization constant.

For $N=3$,  $(D^{(3)}-D^{(2)})=1-1=0$, and therefore there is no 
$SL^{(1)}(2,R)$ vacuum.

For $N=4$, the only vacuum and the corresponding excited states are:
\be
\mid N=4,n\rangle=C_{4,n}
(\hat{L}_{-1})^n(-\frac{3}{5}\hat{L}_{-4}+\hat{L}_{-2}\hat{L}_{-2})
\mid\!0\rangle.
\ee

For $N=5$, as for $N=3$,  there is no vacuum. 

For $N=6$, we have the following vacua (chosen as orthogonal):
\bea    
\mid N=6,n=0,1\rangle&=&C_{6,0,1}
(\frac{-1}{7}\hat{L}_{-6}-\frac{8}{5}\hat{L}_{-4}\hat{L}_{-2}+\hat{L}_{-3}
\hat{L}_{-3})\mid\!0\rangle
\hbox{\ and \ } \\
\mid N=6,n=0,2\rangle&=&C_{6,0,2}
(\frac{_{1183-6080c}}{^{-18646+20160c}}\hat{L}_{-6}+\frac{_{28013+211680c}}
{^{74584-80640c}}\hat{L}_{-4}\hat{L}_{-2} \nn \\
&+&\hat{L}_3\hat{L}_3+\frac{_{10365+22400c}}{^{-18646+20160c}}
\hat{L}_{-2}\hat{L}_{-2}\hat{L}_{-2})\mid\!0\rangle.\nn
\eea
which generate the corresponding representations.

Now, with each irreducible representation of $SL^{(1)}(2,R)$ we 
associate a space-time geometry as the support of the $C^*$-algebra
generated by the corresponding carrier space. This construction can be made 
in general through the Gelfand-Kolmogorov theory \cite{GK,Connes}.

If we considered the case $c\leq 1$  
(see (\ref{irred}) for which the standard Verma module approach leads
to the existence of null-vector states or, equivalently in our scheme,
when the carrier space
of the representation is the solution to a higher-order polarization 
\cite{Navarro-Salas}), we should take into account that the $C^*$-algebra 
constructed from these wave functions 
would  not be a subalgebra of the space of functions on the group 
(it would not even be commutative) and the general Gelfand-Naimark theory 
\cite{GN} should be used to recover a geometry, which would prove to
be non-commutative. Here, we do not undertake the analysis of this interesting 
case (although we shall do so in the near future), and consider only the 
simpler
representations in which no higher-order polarizations are required so that 
no non-commutative geometry emerges. 
In these particular cases ($c>1$), the process of finding the 
support 
space for each $C^*$-algebra generated by a $SL^{(1)}(2,R)$ irreducible 
representation is not 
involved to a great degree. In fact, we have only to realize that from a given 
$SL^{(1)}(2,R)$ 
irreducible representation $R^{(N)}$, a basis for the complex functions on
the hyperboloid (homogeneous space of $SL(2,R)$) can be obtained from the 
reduction of the tensorial products of $R^{(N)}$ via the 
Clebsch-Gordan series. In this way, we recover an AdS space-time, which is the 
homogeneous space associated with the highest-weight representations of 
$SL(2,R)$\footnote{In fact, a particle moving on AdS space-time is a physical 
system whose quantum space of solutions (Hilbert space) is the same of our 
$SL^{(1)}(2,R)$ representations, and which has an AdS  
space as the configuration space \cite{Bisquert,Manolo}. This allows us to 
identify AdS as the space-time associated with the $SL^{(1)}(2,R)$ 
representations we have found.} 
, the ones appearing in the Virasoro reduction (dS is linked 
to non-highest-weight representations).
Thus, for each $SL^{(1)}(2,R)$ representation, we have a space-time and,
therefore, we find a collection of space-times which are realized 
simultaneously in the theory.

\subsection{Physical interpretation}

Before providing a physical interpretation of this model, let us assign 
dimensions to the objects appearing in it. 
A glance at the commutation relations of Virasoro algebra 
(\ref{conmutacion}), reveals that the integers appearing in it have the same 
dimension
as the generators, dimensions which can be determined if we identify 
(classically) the 
parameters of the group as space-time variables (see classical motion 
equations (\ref{motion})).  Thus, the dimension of generators and integers
 is $(Length)^{-1}$.
From this, we conclude that $c$ has dimensions of $(Length)$ 
and $c'$ of $(Length)^{-1}$.

It is important to redefine the integers as being intrinsically 
dimensionless (if not, we cannot make a physical analysis of mathematically
well-defined expressions such as $\frac{c'}{c}=r^2$ in (\ref{r}),
because we do not have a scale to determine whether an integer is large
or small).
Therefore, we should introduce a $(Length)^{-1}$ dimensional constant $a$ and 
redefine $n\rightarrow\frac 1{a}n$ in all the expressions in the text (we 
have not done so from the very beginning in order not to create confusion with
the existing literature on the Virasoro algebra).
  
We have encountered three fundamental distances in our model: $c$, 
$\frac 1{c'}$ and $a$. In the critical case in which space-time appears,
there is a relationship between the distances ($c=c'a^2$, the dimensionally 
correct version
of $c=c'$) and we have only two  independent ones ($c$ and $a$, for instance).
One of these is related to the notion of long distance 
in the space-time model (the {\it radius}), the other with a short one. From 
arguments to be presented below, we associate $c$ with the long one and 
$a$ with the short one, while the role of the Planck constant is played by
$\frac{a}{c}$.

The constant $\frac{a}{c}$ can be used to redefine the generators in the theory
as is usual in Quantum Mechanics:
\be
\hat{H}_n=\frac{a}{c}\hat{L}_n \; .
\ee
We physically interpret each vector in a $SL^{(1)}(2,R)$ representation of 
index $N$ as a state
of the whole space-time defined by this representation. These 
states are eigenvectors of the kinematical operator $\hat{H}_0$, which can be 
interpreted as the energy \footnote{In fact, the expression of $L_0$ in terms
of the basic variables $L_n \; (\mid n\mid \geq 2)$ (\ref{invNoet}) is a 
generalization of the harmonic oscillator energy and parallels the classical
version of the Sugawara construction of the Hamiltonian in Conformal Field 
Theory \cite{Sugaw}.}.
Thus, the maximal-weight vector of the 
representation, $\mid\! N, 0\rangle$,
is the fundamental state of the corresponding space-time, while the action 
of $\hat{H}_{-1}$ moves
space-time to excited states:
\bea
\hat{H}_0((\hat{H}_{-1})^n\mid N, 0\rangle)&=&\frac{a}{c}\frac{(N+n)}{a}
(\hat{H}_{-1})^n\mid N, 0\rangle \\
Energy(n)&=&\frac{N+n}{c}. \nn
\eea
The vacuum of the Virasoro representation, $\mid\! 0 \rangle$, is interpreted 
as the physical vacuum of the (whole) Universe \footnote{We
use the term ``Universe'' in referring to the entire Virasoro representation 
(the entire physical system), and ``space-time'' referring to the geometry  
related to $SL^{(1)}(2,R)$ representations.} in which we do not even have 
a space-time (is the 
trivial representation of $SL^{(1)}(2,R)$). The energy of this vacuum is $0$,
as it should be, but the reason is by no means trivial: it is just 
a consequence of $c$ and $c'$ being in the critical value $c=c'$.
   
We have been using the term space-time in the text, while this is not quite 
precise, as we have no metric notion yet. The reconstruction from the 
$C^*$-algebra does not provide a metric. The only primary metric we can 
consider 
in the context of our model is the one induced on the hyperboloid
from the Killing metric of $SL^{(1)}(2,R)$, which turns out to be AdS metric
(as we said before) due the the presence of highest-weight representations. 
To implement the constraint which allows us
to induce this metric, we have to give the {\it radius} of the hyperboloid. We 
search in the model for a distance notion which should be completely 
characterised by the Virasoro representation (i.e. by $c$) and by the 
$SL^{(1)}(2,R)$ representation ($N$). A length that fulfills these 
requirements is
the Casimir in terms of $\hat{H}_0, \hat{H}_1, \hat{H}_{-1}$:
\be
\frac{1}{R^2}\equiv {\hat{H}_0}^2-\frac {1}{2}(\hat{H}_1\hat{H}_{-1}+
\hat{H}_{-1}\hat{H}_1)=
(\frac{a}{c})^2 \frac{N(N-1)}{a^2}= \frac{N(N-1)}{c^2}.
\ee
We have an AdS metric on the hyperboloid given by:
\bea
ds^2={dv}^2+{d\bar{\lambda}}^2-{du}^2\\
v^2+\bar{\lambda}^2-u^2=\frac{c^2}{N(N-1)} \; , \nn
\eea
where $u$ and $v$ are linear combinations of $l^1$ and $l^{-1}$, which make
the corresponding  momentum generators hermitian.
Therefore, we have an AdS space-time support.

Up to now, we have been concerned with the $SL^{(1)}(2,R)$ symmetry, which  
provides an 
ensemble of AdS space-times with {\it radii} $\frac{c}{\sqrt{N(N-1)}}$ 
associated with each $SL^{(1)}(2,R)$ representation. On them the 
$\hat{H}_{-1}$ operator 
acts by creating excited states of these space-times. No relationships among 
the different $SL(2,R)$ representations have been reported. 
Let us now consider the $\hat{H}_n$  ($\mid n\mid\geq 2$) 
gravitational modes. As they do not preserve the $SL^{(1)}(2,R)$ 
representations, 
they have the effect of transforming a state of a definite space-time, into a 
linear combination of states of different space-times. That is, if we start 
from a state of a space-time of {\it radius} $R$, after the
action of gravity the state that describes space-time is spread over 
space-times of different {\it radii}. Taking advantage of the orthogonality
of $SL^{(1)}(2,R)$ subrepresentations, the probability for a  
state 
to have a definite {\it radius}, can be computed by simply using the 
orthogonal projector on the 
appropriate $SL^{(1)}(2,R)$ representation.

This is the essence of our quantum-gravity model: the Universe is not just a 
space-time (a $SL^{(1)}(2,R)$ representation), but the whole ensemble of them.
A state of the Universe is a superposition of space-times with different radii 
(states in different $SL^{(1)}(2,R)$ representations). We 
cannot speak of the {\it radius} of the Universe; only the probability that 
the 
Universe has a certain {\it radius} makes real sense. The effect of gravity is 
that of changing the 
{\it radii} distribution of the Universe ($\hat{H}_{n \geq 2}$ move 
the distribution
towards smaller {\it radii}, while $\hat{H}_{n \leq 2}$ bring about larger 
{\it radii}) and, on a
specific space-time, producing linear excitations ($\hat{H}_{\mid n\mid=1}$) 
which eventually might be interpreted as quantum states of a free ``particle''
of mass $m=m(N)$ moving on this AdS space-time \footnote{A specific 
determination of
$m$ could be made by comparing the present $SL(2,R)$ states with those found 
in \cite{Fronsdal,Isham,Bisquert}. In fact, the wave functions in those papers
(when restricted to the $(1+1)$-dimensional case) support an irreducible 
representation of $SL(2,R)$ with index $N=\frac{cR}{m\hbar}$, where in this
expression $c$ is the
speed of light, $\hbar$ the standard Planck constant and $R$ the {\it radius}
of the AdS space-time.}. 

It should be stressed that, since we are dealing with maximal-weight 
representations, the net effect of the gravitational modes is
the decreasing of the average {\it radius} ($\hat{H}_{n\geq 2}$ eventually 
annhilate a given state of the Universe, while $\hat{H}_{n\leq 2}$ do not). 

A remarkable property of the underlying symmetry, the Virasoro group, is that
(as pointed out in the {\bf Introduction}) it can be realized as the 
diffeomorphism
group of a given manifold ($S^1$). Thus, the quantum operators of the theory 
can be thought of as being the quantum version of (non-linear or general)
changes of reference, traditionally considered as gauge transformations. In
the present model, the Virasoro (quantum) operators generate true dynamical 
changes in the sense that they have a non-trivial action on the Hilbert space 
(quantum solution manifold). For instance, the operator $\hat{H}_2$ takes the 
state
$\mid N=4,n=0 \rangle\equiv (-\frac{3}{5}\hat{L}_{-4}+\hat{L}_{-2}
\hat{L}_{-2})\mid\! 0\rangle$,
representing a ground AdS space-time of {\it radius} $\frac{c}
{\sqrt{3\cdot 4}}$ to 
another ground AdS space-time, but this time of {\it radius} 
$\frac{c}{\sqrt{1\cdot 2}}$. Only for $c\rightarrow\infty$, the classical 
limit (see next subsection), this transformation can be considered as a gauge 
transformation. 

In fact, at this limit, we find that the energy of the ground
state goes to zero
\linebreak
($Energy\mid\!N,0\rangle\rightarrow 0$) and the 
{\it radius} to infinity ($R=\frac{c}{\sqrt{N(N-1)}}\rightarrow\infty$) for
all the space-times. Therefore, they are physically indistinguishable
and it makes sense to identify them, resulting in the existence of a single 
space-time in the classical limit $c\rightarrow\infty$. This implies the loss
of dynamical content of the $\hat{L}_n$ modes, which act as gauge 
transformations
in that single space. We recover the gauge nature of the diffeomorphism
but only in the classical limit. The solution manifold under the 
diffeomorphism constraints would go to a one-degree of freedom phase-space,
one $q$ and one $p$ (which is formally equivalent
to that of a single particle moving in a fixed space-time).
This is a rather standard situation in other approaches to 
2D-gravity, where the diffeomorphism constraints are imposed prior to the 
quantization \cite{ligaduras}.

\subsection{The classical limit}
Finally, let us consider the (semi)-classical limit of the model. 
The main interest of this limit is really the justification of the statements
made about the different constants which  
appeared in the previous subsection.

It can be argued \cite{Witten}, using the Virasoro
Poisson brackets (in the original form (\ref{conmutacion})), that the 
semiclassical region for the quantization of the Virasoro group corresponds 
to large values of the true cohomology parameter $c$. The Planck constant
proves to be 
$\sim \frac 1{c}$ ($\frac {a}{c}$ when the dimensional constant $a$ is 
introduced);  that is, in the semiclassical region, the fundamental distance 
$c$ is much larger than $a$.

Consistency with the classical limit is the reason for chosing $c$ as being 
related to the large fundamental distance (and eventually to the Universe 
{\it radius}) and $a$ to the small one. The condition that characterizes the 
class of Virasoro representations under study (i.e. $c>1$) prevents the long 
distance $c$ from getting 
smaller than
the short length $a$. In fact, it imposes $\frac{c}{a}>1$ (the dimensionally
correct version of ($c>1$), so that we always
have $c>a$). Long and short fundamental distances are, in this way, 
well-defined notions in the sense that they do not cross each other. 
This is no 
longer valid, however, for the severe quantum region, $c<1$.

The {\it radius} is thus $\frac{c}{\sqrt{N(N-1)}}$.  Therefore, a 
semi-classical region of the system (large
$c$) corresponds to a large value of the {\it radius} $R$ of our 
space-time support. With respect to the metric on the hyperboloid,
this imposes that $\mid K\mid\ll 1$, so that we approach a Minkowski space-time
\footnote{In loose terms (we repeat that there is
no cosmological constant in two dimensions), the ``cosmological constant'' 
$\Lambda$ ($\sim K$) goes to zero in the semi-classical region.}.

Let us develop the classical limit in more detail and then compare it with the 
classical formalism described at the end of {\bf Section 2}, in order the 
identify the physical content of the theory at this phenomenological limit.

At the quantum level the dynamics is described by the action of the modes
$\hat{L}_n\;(\mid\! n\!\mid\geq2)$, the effect of which is that of mixing
states
of space-times with (in general) different radii, thus affecting a quantum 
notion of {\it distance}. In the configuration-space description the dynamics
can be encoded in the field operator $\hat{F}$, obtained from the expression 
of $F(\lambda,u,v)$ in (\ref{F}) by replacing the constants ${\cal L}_n$ (or
equivalently $L_n$'s) with the corresponding quantum operators $\hat{L}_n$.
At the classical limit, where there is a single space-time (as explained at 
the end of the previous section), the phenomenological result of the 
dynamical transformations must be relegated to that of producing changes
in the classical {\it distance}. In a classical theory, the object that models 
such changes in the {\it distance} is a dynamical metric field. Thus, in the 
limit $c\rightarrow\infty$ of this theory, we expect the field 
$F(\lambda,u,v)$ to be
associated with the dynamical part of a metric.

More precisely, the metric tensor must adopt the form
\bea
g^{\mu\nu}={\eta^{\mu\nu}}\;+\;{g^{\mu\nu}}_{dyn} 
\eea
\ni where ${\eta^{\mu\nu}}$ is the background metric inherited from the rigid 
AdS metric of each of the coexisting space-times in the quantum theory (and it 
is associated with the kinematical degrees of freedom $L_0,L_{\pm 1}$) and
${g^{\mu\nu}}_{dyn}$ is the dynamical part, which must be derived in terms of 
the classical field $F(\lambda,u,v)$.

To determine the explicit form of the metric
${g^{\mu\nu}}_{dyn}$,
we resort to the classical formulation developed in the last part of
{\bf Section 2}.

Firstly, we constrain the $SL(2,R)$ parameters $\lambda,u,v$ in $F$, in order 
to fall down to an AdS space-time of radius $R$ (this is the classical
analogue of the restriction in the quantum theory to an $SL(2,R)$ 
representation by imposing the Casimir 
constraint). Thus, from the expression obtained in {\bf Section 2},
\bea
F(u,v,\lambda)=\lambda+\sum_{n\neq0}l_n(u,v)e^{in\lambda}\;,  
\eea
\ni we obtain
\bea
F_{AdS}(u,\lambda)&\equiv& \int dv\; \delta(v^2+\bar{\lambda}^2-u^2-R^2)
F(u,v,\lambda)=\nn \\
&=&\int dv\; \delta(v^2+\bar{\lambda}^2-u^2-R^2)(\lambda+\sum_{n\neq0}l_n(u,v)
e^{in\lambda})\; .\nn 
\eea
This constraint forces $v$ to be of the form
\bea
v=\sqrt{R^2 -\hat{\lambda}^2+u^2}=R\sqrt{1-(\frac{\bar{\lambda}}{R})^2+
(\frac{u}{R})^2}\; , 
\eea
\ni which in the classical limit, $R\rightarrow\infty$, simplifies to
$v\sim R$. Therefore:
\bea
l_n(u,v)&\sim& l_n(u,R\rightarrow\infty)=l_n(u)  \\
F^{R\rightarrow\infty}_{AdS}(u,\lambda)&=&\lambda+\sum_{n\neq0}l_n(u)
e^{in\lambda}\equiv F(u,\lambda) \nn 
\eea
This expression can be directly inverted and the 
expression for the $l_n$ of reference \cite{Miguel}, as Fourier coefficients
of the diffeomorphisms of $S^1$, is recovered. From this, and the explicit
form of $\Theta$, we get the expression for the action of the field $F(u,
\lambda)$:
\be 
S=\int\Theta=-\frac {c}{48\pi}(\int dud\lambda \frac {\partial_{u} F}{\partial_
{\lambda} F}
(\frac {\partial_{\lambda}^3 F}{\partial_{\lambda} F} - 2 \frac{(\partial_
{\lambda}^2 F)^2}
{(\partial_{\lambda} F)^2}) + \int dud\lambda \frac {\partial_u F}{\partial_
\lambda F}( \partial_{\lambda} F - 1))\;\; . \label{S}
\ee
At this point, we recognize the form of Polyakov's action (\ref{Pol}) with a 
corrective term. The role of the light-cone variables $x^-$ and $x^+$ is
played by $\lambda$ and $u$, respectively. Repeating in reverse order the 
arguments
of \cite{Polyakov}, we identify the previous expression with the action of a 
dynamical metric of the form,
\bea
ds^2_{dyn}= \partial_\lambda F d\lambda du \;, 
\eea
which arises as linked to the conformal anomaly \footnote{The conformal 
anomaly 
is present in our model, as can be computed from the Noether invariants $L_n$
with the Poisson bracket derived from $d\Theta$, or directly from the 
commutators of the quantum operators $\hat{L}_n$. Note that if $c=0$, the 
action vanishes.}. 

Thus, the complete form of the metric on the space-time at the classical limit
is:
\bea
ds^2=ds^2_{AdS(R\gg 1)} \;+\; \partial_\lambda F d\lambda du \;. 
\eea
We see that, due to the presence of the background term, the nature of 
$\lambda$ and $u$ is no longer that of light-cone variables, but rather 
of time and 
space, as dictated from the AdS metric. Only in the regime 
($c\rightarrow\infty$, $\mid\partial_\lambda F\mid\gg1$) can the background 
term be neglected and can we properly recover the corrected Polyakov action. 
The
corrective term has already been found in the literature \cite{W}, 
where it was interpreted as being related to an outer field
U, whereas here it is crucial for the consistency of space-time.

\section{Conclusions}

We have reviewed the Virasoro group as the basic symmetry of a model for 
two-dimensional quantum gravity (without matter), 
avoiding the assumption of the existence of external parameters which build the
space-time manifold. In this context, we have seen that such space-time emerges
only for the critical value of the anomaly $c=c'$, as a consequence of
the fundamental role played by cohomology (and pseudo-cohomology) in the 
determination of the dynamical content of the degrees of freedom of a theory.
Nevertheless, a well-defined theory out of this critical value of the 
extension does exist (we have an explicit realization of the algebra of 
operators) and we
can argue that even in those cases in which the notion of space-time makes no
sense, we have
a ``physical'' system which evolves according to some proper time.

If we insist on the notion of space-time, non-commutative 
geometry ideas are well suited for the implementation of this notion. In fact,
non-commutative $C^*$-algebras leading to non-commutative geometries can 
occur if higher-order polarizations are needed to reduce the representations.
In generalizations of this model one must be prepared to 
deal with non-commutative geometry.

The notion of space-time in our approach is rather unusual, if compared with 
other schemes,
in some respects here summarized:
\begin{itemize}
\item It appears only for a critical value of the central extensions of the 
group.
\item For a given value of $c=c'$, it is a superposition of standard 
space-times 
with different {\it radii} (the different representations of $SL(2,R)$ that
appear in the Virasoro representation), with a weight given by the 
degeneration factors.
\item The quantum analogues of general changes of variables are not 
necessarily 
gauge transformations. General covariance may be properly realized in the 
classical limit. It reinforces the idea that diffeomorphism constraints
should not be fully imposed prior to quantization.

\item We have found a relationship between  two fundamental constants, the 
curvature $K$ (related with the {\it radius} $R$ of the Universe) 
and the Planck constant $\hbar$ (related to $c$ \footnote{Note that 
this is an ``interaction'' Planck constant in 
the sense that it is essentially related to gravity. Here it is important
with respect to the classical limit.}; see \cite{Witten}):\\
\be
K \sim R^{-2}\sim c^{-2}\sim \hbar^2.
\ee
Thus, if we look at the Universe in a classical way ($h\rightarrow 0$)
we find that the curvature goes to zero.
\end {itemize}
Although the fundamental goal of the present paper was 
to clarify the way space-time notion emerges, the introduction of matter in 
the model should be studied next. This can be 
accomplished by considering the semi-direct action of Virasoro on a Kac-Moody
group. Support for this idea can be found in \cite{Nicolai}, where, by the
use of a 
completely different approach, the structure of the solution-space manifold 
for 2D gravity with matter is identified as a $\frac {W\otimes_sG^\infty}
{K\otimes_sH^\infty}$ homogeneous space (something expected in the 
quantization of a Kac-Moody group with a Virasoro semidirect action). 
The important point is that the 
present work suggests the separation of 
the problem
of space-time from that of matter in the $W\otimes_sG^\infty$ 
quantization.

Another unavoidable question is that we have not dealt with Einstenian 
gravity, 
but rather with a higher-order correction to it.
In two dimensions, classical Einstein gravity is trivial, but 
in going from $1+1$ to $3+1$ dimensions, we should find an analogue
of the Virasoro group and,
in addition, a precise framework through which 
Einstenian (or a quantum version of it) enters the scene.
Also, going to higher dimensions within the present scheme opens the 
possibility
of having natural transitions between space-times with different topologies as
homogeneous spaces associated with a characteristic subgroup larger than 
$SL(2,R)$.

A final general remark is that in GAQ any generator in the characteristic 
subalgebra can
be written as a function of the dynamical ones (that is, the basic ones). In
our model this means that space-time generators are expressed in terms of
quantum gravity operators: space-time is thus constructed from interaction.

\section{Appendix A}

\begin{itemize}
\item  Reduction of ${\cal H}_{(c,c)}$ under $SL^{(1)}(2,R)$.

{\it i)} $SL^{(1)}(2,R)$ maximal-weight vectors have definite a level.

{\it Proof}: let $\mid m \rangle$ be in an irreducible representation of 
$SL^{(1)}(2,R)$, satisfying $\hat{L}_1\mid~m\rangle~=~0$ (maximal-weight 
vector,
which is unique in the representation). Let us consider the vector 
$\hat{L}_0\mid m \rangle$, which is in the same representation that 
$\mid m \rangle$. Then\\
\be
\hat{L}_1(\hat{L}_0\mid m \rangle)= (\hat{L}_1 + \hat{L}_0\hat{L}_1)\mid m
\rangle=0 \;\; .
\ee
That is, $\hat{L}_0\mid m \rangle$ is also a maximal weight vector, which 
implies:\\
\be
\hat{L}_0\mid m\rangle=N\mid m\rangle \;\; \Rightarrow \;\; 
\mid m\rangle \;\;\hbox{eigenvector of } L_0.
\ee
Furthermore, the value of the Casimir on $\mid m\rangle$ (and in all the 
representation) is $N(N-1)$:\\
\be
({\hat{L}_0}^2-\frac 1{2}(\hat{L}_1\hat{L}_{-1}+\hat{L}_{-1}\hat{L}_{1}))\mid 
m\rangle=
({\hat{L}_0}^2-\hat{L}_0)\mid m\rangle=(N^2-N)\mid m\rangle=N(N-1)\mid m\rangle
\ee
\\
\ni{\it ii)} Vectors inside an irreducible representation have a level higher
than the level of their maximal-weight vector.

{\it Proof}: Directly from the construction of the vectors 
${(\hat{L}_{-1})}^n \mid m\rangle$.

{\it iii)} In the level $N$ there are $D^{(N)}-D^{(N-1)}$ maximal-weight 
vectors, where $D^{(N)}$ is the dimension of the level $N$.

{\it Proof}: We use induction on N.

\ni For N=2  ($D^{(2)}-D^{(1)}=1-0=1$), and in fact the only independent
vector in the $N=2$ level, $\hat{L}_{-2}\mid\! 0\rangle$, is a maximal-weight 
vector:
$\hat{L}_1\hat{L}_{-2}\mid\! 0\rangle=0$ (we can also check the validity of 
our assertion
for $N=3$, $D^{(3)}-D^{(2)}= 1-1=0$, or $N=4$, $D^{(4)}-D^{(3)}= 2-1=1$).

Assuming it for $N-1$, let us consider the level $N$. There are $D^{(N)}$ 
independent vectors, $D^{(N-1)}$ of which belong to representations induced
from level $N-1$ by the action of $L_{-1}$. Therefore we can find $D^{(N)}-
D^{(N-1)}$ independent vectors which do not belong to representations 
constructed from maximal-weight vectors of a lower level and by {\it ii)} they
can neither be obtained from maximal-weight vectors of higher level. Thus, as
belonging to some irreducible representation (at this point we are assuming
complete reducibility), they have to be maximal-weight
vectors themselves.
\ni They generate the $D^{(N)}-D^{(N-1)}$ irreducible representations of
$N(N-1)$ Casimir (by {\it i)}).

(Note: {\it ii)} and {\it iii)} can be directly bypassed, noting that the
homomorphism $\hat{L}_{-1}$ from level $N$ to level $N-1$ is a surjective one.
Thus, we use
\be
dim(Im\; L_{-1})+dim(Ker\; L_{-1})= dim( Level\;N) 
\ee
to achieve the desired result, without assuming complete reducibility, but
attaining it in a constructive way).
 
Finally we have,\\
\be
{\cal H}_{(c,c)}=\bigoplus_{N} (D^{(N)}-D^{(N-1)})R^{(N)} 
\ee

\item Orthogonality of the $SL(2,R)$ representations.

{\it i)} Different Virasoro levels are orthogonal.

Let us consider a vector $\hat{L}_{n_j}...\hat{L}_{n_1}\mid\! 0\rangle$ 
on level
$N$ and
the vector $\hat{L}_{m_j}...\hat{L}_{m_1}\mid\! 0\rangle$ on level $M<N$. When 
we construct
the scalar product, $\langle 0\mid \hat{L}_{-m_1}...\hat{L}_{-m_j}
\hat{L}_{n_j}...\hat{L}_{n_1}\mid 
0\rangle$, we observe that the vector $\hat{L}_{-m_1}...\hat{L}_{-m_j}
\hat{L}_{n_j}...\hat{L}_{n_1}
\mid\! 0\rangle$ belongs to $N-M$ level, and can be written as a linear 
combination of a basis of that level. Each element of the basis annhilate
$\langle 0\mid$ by the polarization conditions.

{\it ii)} States of the same level in different $SL(2,R)$ representations 
are orthogonal. 
 
Let us consider two maximal weight states $\mid N_1 \rangle$ and 
$\mid N_2 \rangle$, corresponding to different representations of level $N_1$ 
and $N_2$ ($N_1\leq N_2$), respectively. Now, let us consider the scalar 
product of two 
states $(\hat{L}_{-1})^{n_1}\mid N_1\rangle$ and $(\hat{L}_{-1})^{n_2}
\mid N_2\rangle$, such that $n_1+N_1=n_2+N_2$:
\be
\langle N_1 \mid (\hat{L}_{1})^{n_1}(\hat{L}_{-1})^{n_2}\mid N_2\rangle=
\langle N_1 \mid (\hat{L}_{1})^{n_1-n_2}(\hat{L}_{1})^{n_2}
(\hat{L}_{-1})^{n_2}\mid N_2\rangle 
\ee
The operator $(\hat{L}_{1})^{n_2}(\hat{L}_{-1})^{n_2}$ can always be written 
in the form $(...)L_1 + L_0$. The first term directly annihilates the vector 
$\mid N_2\rangle$,
while $\mid N_2\rangle$ is an eigenvector of $\hat{L}_0$. Thus,\\
\ni a) For $n_1-n_2>0$, $(\hat{L}_{1})^{n_1-n_2}$ annihilates 
$\mid N_2\rangle$.\\
\ni b) For $n_1=n_2$, we can always choose $\mid N_1\rangle$ orthogonal to 
$\mid N_2\rangle$.

We have proven the orthogonality of the two vectors. 

\end{itemize}

\section{ Appendix B}
\begin{itemize}
\item Expression for invariant vector fields:\\
\bea
\tilde{X}_{l^k}^R&=&\frac{\!\!\partial} {\partial l^k}+ikl^{m-k}
\frac{\!\!\partial}
{\partial l^m}+\frac{\left( ik\right) ^2}{2!}l^nl^{m-n-k}\frac \partial
{\partial l^m}+...+\frac{\left( ik\right) ^j}{j!}\sum
\limits_{n_1+...+n_j=m-k}\!\!\!l^{n_1}...l^{n_j}\frac{\!\!\partial}
{\partial l^m}\nn \\
&&+...+\frac c{24}\{\left( -i\right) k^2\left( -k\right) l^{-k}+\frac{\left(
-i\right) ^2}{2!}k^2\sum\limits_{n_1+n_2=-k}\!\!\!\left(
n_1^2+n_2^2+n_1n_2\right) l^{n_1}l^{n_2}+...+\nn \\
&&\frac{\left( -i\right) ^j}{j!}k^2\sum\limits_{n_1+...+n_j=-k}P^{\left(
j\right) }\left( n_1,...,n_j\right) l^{n_1}...l^{n_j}+...\}\Xi - \\
&&\frac{c}{24}\{ikl^{-k}+\frac{\left( ik\right) ^2}{2!}l^{n_1}l^{-k-n_1}+
...+\frac{\left( ik\right) ^j}{j!}\sum
\limits_{n_1+...+n_j=-k}l^{n_1}...l^{n_j}+...\}\Xi \nn \\ \label{Camposiz}
\tilde{X}_\zeta ^R&=&i\zeta \frac \partial {\partial \zeta }\equiv \Xi \; ,\nn
\eea

\ni and\\

\bea
\tilde{X}_{l^k}^L&=&\frac \partial {\partial l^k}+i\left( m-k\right)
l^{m-k}\frac \partial {\partial l^m}-\frac{c^2}{24}k \label{XRlk} \nn \\
&&\left\{ \left( -i\right) \left( -k\right) l^{-k}+...+\frac{\left( -i\right)
^j}j\sum\limits_{n_1+...+n_j=-m}n_1...n_jl^{n_1}...l^{n_j}+...\right\} \Xi  \\
&&-\frac{c^{\prime }}{24}i\left( -k\right) l^{-k}\Xi \nn
\eea
\\

\end{itemize}

\section{Acknowledgements}

We thank V.G. Kac for discussions about some mathematical aspects of the 
paper and M. Calixto, J. Guerrero and G. Mena for useful comments and for
reading the manuscript. We also thank ESI (Vienna) for its hospitality at the 
early stage of the work.

\end{document}